\documentclass[twocolumn, switch]{article} 

\usepackage{preprint}

\usepackage{amsmath, amsthm, amssymb, amsfonts}

\usepackage[backend=biber,style=numeric-comp,isbn=false]{biblatex} 

\usepackage[utf8]{inputenc}	
\usepackage[T1]{fontenc}	
\usepackage{xcolor}		
\usepackage[colorlinks = true,
            linkcolor = purple,
            urlcolor  = blue,
            citecolor = cyan,
            anchorcolor = black]{hyperref}	
\usepackage{booktabs} 		
\usepackage{multirow}		
\usepackage{nicefrac}		
\usepackage{microtype}		
\usepackage{lineno}		
\usepackage{float}			

\usepackage{lipsum}		

\usepackage[per-mode=symbol,separate-uncertainty=true,multi-part-units = single,list-units=single,product-units = power]{siunitx}	
\DeclareSIUnit{\wtpercent}{wt\%}
\DeclareSIUnit\px{px}
\DeclareSIUnit\mercury{Hg}

\usepackage{graphics}

\usepackage{newfloat}
\DeclareFloatingEnvironment[name={Supplementary Figure}]{suppfigure}
\usepackage{sidecap}
\sidecaptionvpos{figure}{c}

\usepackage{titlesec}
\titlespacing\section{0pt}{12pt plus 3pt minus 3pt}{1pt plus 1pt minus 1pt}
\titlespacing\subsection{0pt}{10pt plus 3pt minus 3pt}{1pt plus 1pt minus 1pt}
\titlespacing\subsubsection{0pt}{8pt plus 3pt minus 3pt}{1pt plus 1pt minus 1pt}

\title{Experimental analysis of a Multilayer Flow Modulator deployment in an AAA with incorporated branch}

\usepackage{authblk}

\author[1*]{Simon Tupin}
\author[2]{Kei Takase}
\author[1]{Makoto Ohta}
\affil[1]{Biomedical Flow Dynamics Laboratory, Institute of Fluid Science, Tohoku University, Sendai, 980-8577 Miyagi, Japan}
\affil[2]{Department of Diagnostic Radiology, Tohoku University School of Medicine, Sendai, 980-8575 Miyagi, Japan}
\affil[*]{{correspondence:} \textsf{s.tupin@tohoku.ac.jp} \href{https://orcid.org/0000-0003-0982-8210}{0000-0003-0982-8210}}

\begin{document}

\twocolumn[ 
  \begin{@twocolumnfalse} 
  
\maketitle

\begin{abstract}
\textbf{Purpose}: 
To experimentally evaluate the effect of a Multilayer Flow Modulator (MFM) device deployment in an Abdominal Aortic Aneurysm (AAA) model with and without incorporated branch.\\
\textbf{Methods}: 
An experimental flow and pressure monitoring system was developed to analyze the MFM deployment procedure performed by a qualified radiologist. The pressure and flow rate evolution during the procedure was recorded. Particle Image Velocimetry (PIV) experiments were conducted on models with and without MFM device to evaluate and compare flow patterns and local flow velocity and vorticity in the aneurysm.\\
\textbf{Results}: 
The experiments revealed no significant change in pressure and flow rate during and after deployment of the MFM device. The flow rate of the incorporated branch was fully preserved. On all models, the aneurysmal flow velocity was significantly reduced. In addition, the device modified local flow patterns, reducing vorticity and better feeding the incorporated branch.\\ 
\textbf{Conclusion}: 
This first experimental study provides the basis for a better understanding of the mechanism of the MFM device, which allows intra-aneurysmal flow decrease while preserving incorporated branch and reducing the risk of endoleak type II. The experimental system developed for this study was effective in simulating an endovascular procedure and studying the safety and effectiveness of endovascular devices.

\end{abstract}

\vspace{.5cm}

  \end{@twocolumnfalse}
]

\section{Introduction}

Endovascular aneurysm repair (EVAR) has become the preferred treatment for abdominal aortic aneurysm (AAA) over the invasive open repair technique\parencite{01_Sakalihasan2018}. In this procedure, an endovascular graft component is deployed in the aorta to exclude the aneurysm from the systemic circulation\parencite{02_Wanhainen2019}.

Because of the impermeable nature of the graft, any side branch within the aneurysm will lose its perfusion. One of the major complication of EVAR is the type II endoleak (T2EL) (10-\SI{25}{\percent}\parencite{03_Veith2002}), defined as a reflux from the branch into the aneurysm, which jeopardizes the success of the treatment because the aneurysm may continue to grow and rupture\parencite{01_Sakalihasan2018}. The origin of T2EL is often the Inferior Mesenteric Artery (IMA)\parencite{04_Baum2001}. To avoid T2EL, prior branch embolization\parencite{05_Natrella_2017,06_Bryce2018} or custom chimney graft\parencite{07_Pfister2016} are necessary, increasing operative time, cost and patient burden.

The Multilayer Flow Modulator (MFM, Cardiatis, Isnes, Belgium) is a self-expanding uncovered stent made of cobalt alloy braided wires. This technology excludes the aneurysm by allowing blood to flow through the device to maintain branch patency, while laminating the flow in the aneurysm sac at a lower velocity\parencite{08_Sultan2014}. This flow-diversion technique was first introduced for the treatment of intracranial aneurysms\parencite{09_Murray2019}. Although several animal and \emph{in vivo} studies on the use of the MFM have been published\parencite{10_Sultan2015,11_Benjelloun2016,12_Ibrahim2018}, criticism remains\parencite{13_Mastracci2016,14_Oderich2017,15_Chung2018} and quantitative flow analysis, necessary to understand the mechanism of the device and assess its safety, has not yet been performed\parencite{16_Buck2013}. 

\emph{In vitro} experiments using vascular analogues are often selected for the analysis of endovascular devices\parencite{17_Corbett2008}. These experiments make it possible to measure the evolution of intra-aneurysmal pressure after graft deployment\parencite{18_Xenos2003,19_Chong2003,20_Gawenda2003}. Particle Image Velocimetry (PIV) is also a technique of choice for comparing the effectiveness of endovascular devices, capable of providing both qualitative and quantitative results.\parencite{21_Boersen2017,22_Velde2018}

The purpose of this study was to analyze the effect of the deployment of a MFM device on the intra-aneurysmal flow and the incorporated branch perfusion. Two models representing ideal geometry of an AAA with and without incorporated branch were used to reproduce and record a deployment procedure under pressure and flow rate monitoring. Flow patterns were then evaluated by PIV experiments and used to compare control models with treated ones.

\section{Methods}

\subsection{Model geometry}
\begin{figure*}[p]
	\centering
	\includegraphics[width=0.7\textwidth]{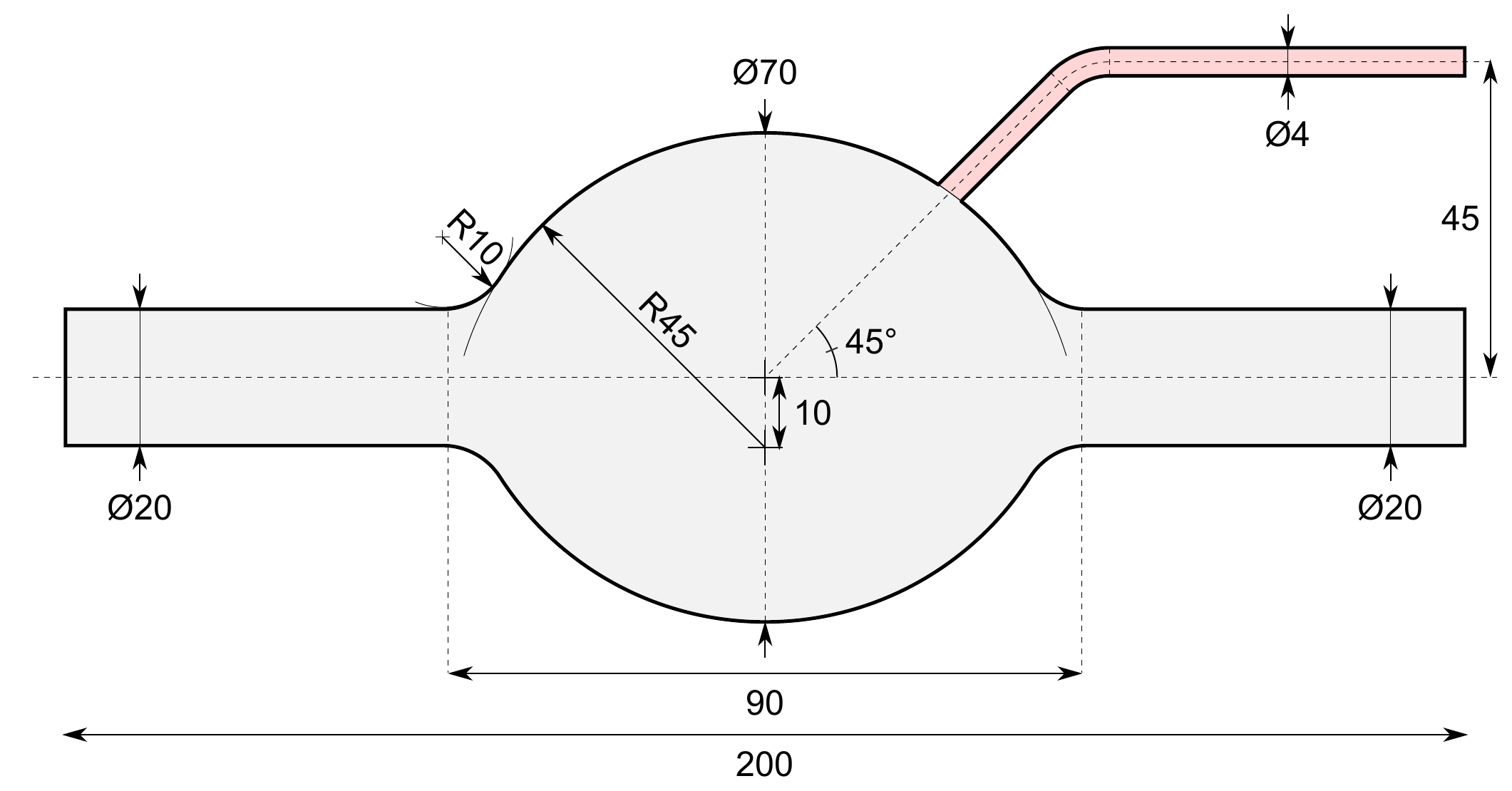}
	\caption{Inner lumen dimensions of the AAA models. Red area denotes the additional branch present in model 2. All dimensions are in mm.}
	\label{fig:fig1}
\end{figure*}

Two ideal symmetrical geometries representing an AAA with and without included IMA were designed as presented in Fig.\ref{fig:fig1}. Silicone box models were manufactured in pairs (R’Tech, Japan) in order to compare two cases: control (untreated) and MFM device deployed model.

\subsection{Circulation system}

\begin{figure*}[p]
	\centering
	\includegraphics[width=\textwidth]{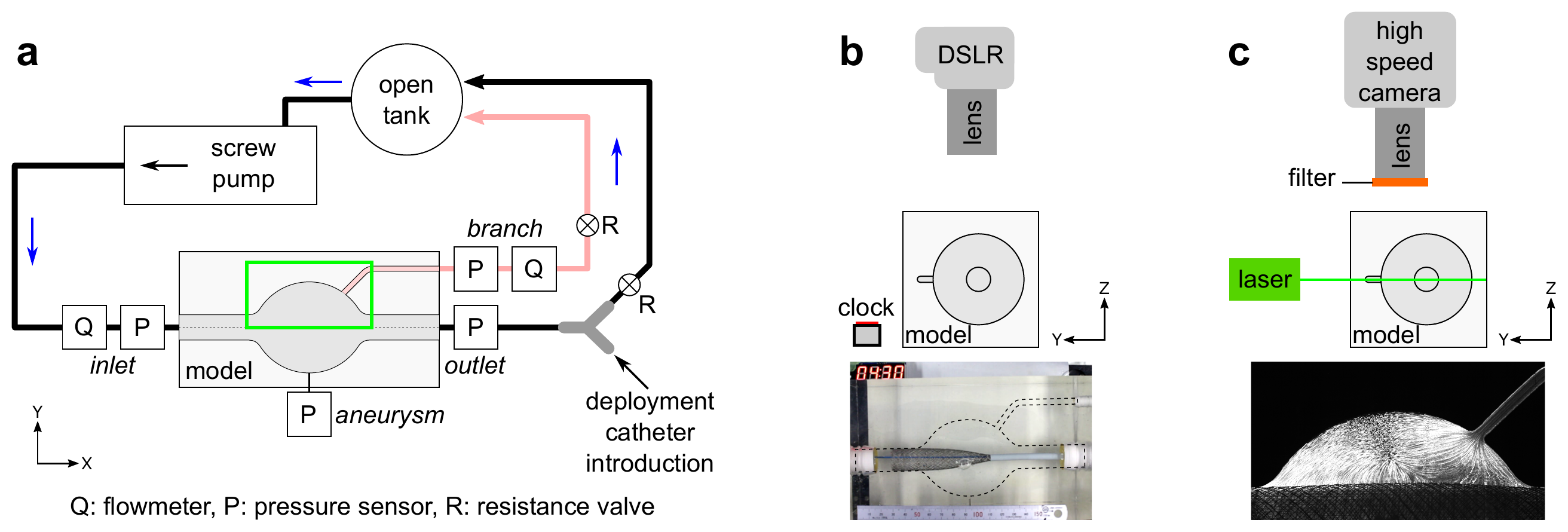}
	\caption{Experimental system. (a) Flow circuit with sensors. Green rectangle denotes the ROI for PIV experiments. Red area denotes the additional branch and sensors present in model 2. (b) Optical set-up for deployment experiment and frame example (model 2). (c) Optical set-up for PIV experiments and particles trace example (model 2).}
	\label{fig:fig2}
\end{figure*}

The experimental system is presented in Fig.\ref{fig:fig2}a. 

A mixture of glycerin, sodium iodide and distilled water was used as blood mimicking fluid (BMF) with properties evaluated at room temperature as follows: a density of \SI{1.2e3}{\kilogram\per\cubic\meter}, a kinematic viscosity of \SI{4.0e-6}{\square\meter\per\second} and a reflective index of \num{1.41}. 

The fluid was stored in an open tank and injected into the silicone model in steady flow condition using an eccentric screw pump (RMP200, R’Tech, Japan).  Boundary conditions were set in the control models as: aortic flow rate of \SI{3}{\liter\per\minute}, branch flow rate of \SI{150}{\milli\liter\per\minute} and pressure level close to \SI{100}{\milli\meter\mercury}.

Pressure sensors (AP-12S, Keyence, Japan) and flow meters (FD-Q20C and FD-SS2A, Keyence, Japan) were used to monitor the flow conditions at each inlet and outlet of the silicone model (Fig.\ref{fig:fig2}a). Pressure inside the aneurysm was recorded using a \SI{0.6}{\milli\meter} diameter fiber optic based pressure sensor (OPP-M250, Opsens Inc., Canada) placed at the edge of the aneurysm through the silicone model wall. Sensors values were recorded at a sampling of \SI{1}{\kilo\hertz} using a CompactDAQ system and a specific software developed on LabVIEW (National Instruments, USA). 

Two valves were used to create resistance in the circuit to set the aortic pressure and the branch flow rate in the control models. Specific connectors were designed in order to easily exchange the silicone model in the experiments without modifying the resistance in the circuit. Observed hemodynamic changes can then be attributed to the MFM device.

\subsection{MFM device and deployment experiments}
A MFM device (Cardiatis, Isnes, Belgium) was deployed in each model. Device sizing and deployment was performed by a qualified radiologist (K.T.) following manufacturer instruction for use (IFU). Selected devices (CTMS25150) measured \O\SI[product-units = single]{25x150}{\milli\meter} and were composed of \num{96} wires of \SI{180}{\micro\meter} diameter.

As during a regular procedure \emph{in vivo}, an introducer sheath (DrySeal Sheath 20Fr, Gore, USA) is used to deploy the MFM device. This equipment was placed in a Y-shaped bifurcation tube attached to the aortic outlet tube to simulate a femoral introduction (Fig.\ref{fig:fig2}a). The clinician first introduced a 0.035-inch guide wire (Radifocus RF-GS35263M, Terumo, Japan) inside the delivery system (\num{18}F~$\times$~\SI{18}{\centi\meter}, Cardiatis, Belgium). This system was then introduced inside the outlet tube through the introducer sheath. The guide wire was advanced beyond the aortic inlet through the aneurysm. The delivery system is then advanced in the model to proceed to the deployment using the “over-the-wire” technique. The MFM device was eventually deployed to bridge the aortic aneurysm by progressively removing the outer sheath of the delivery system while maintaining the position of the proximal end of the MFM more than one centimeter from the entrance of the aneurysm.

Videos of the deployment were recorded and broadcast live to guide the clinician during the deployment, using a DSLR (EOS 100D, Canon, Japan) equipped with a \SI{24}{\milli\meter} lens (Fig.\ref{fig:fig2}b). A digital clock, placed in the field of view of the camera, displayed the time of sensors acquisition to easily synchronize sensors values and images. 

\subsection{PIV experiments}

PIV experiments were conducted to locally image the intra-aneurysmal flow with and without the MFM device. A Nd:YAG solid laser system (BWI-532-300E, B\&W TEK, USA) was used to provide a nominal 1-mm-thick continuous laser sheet through the center plane of the silicone model (Fig.\ref{fig:fig2}c) with a power of \SI{300}{\milli\watt} and a wavelength of 532 nm. Fluorescent microspheres (Fluostar, EBM, Tokyo, Japan) were selected as particles. Images were captured using a high speed camera (Fastcam SA3, Photron, Japan) equipped with a \SI{105}{\milli\meter} lens (Nikon, Japan) and a long pass filter. One thousand and one images were recorded with a resolution of \num{1024 x 512} pixels and \SI{92}{\micro\meter\per\px}. Frame rate was set to \num{60} and \num{1000} fps to image the aneurysmal and incorporated branch flow, respectively. The shutter speed was set to the reciprocal of the frame rate values. 

Recorded images were analyzed using a sum of correlation method with the software DaVis 8.4 (LaVision, Germany). Streamlines, velocity magnitude ($u$) and vorticity ($\omega$) maps were computed using MATLAB 2018b (Mathworks, USA). To quantitatively compare intra-aneurysmal flow, average velocity magnitude ($\overline{u}$) and enstrophy ($\mathcal{E}$) were calculated across the aneurysm domain. The enstrophy ($\mathcal{E}$), defined as the integral of the squared vorticity field, provides a quantification of the amplitude of the vorticity\parencite{23_Deplano2013}.

Flow rate and pressure were also measured during the PIV experiments on each model to quantitatively evaluate the MFM device effect. 

\section{Results}

\subsection{Deployment experiments}
\begin{figure*}[p]
	\centering
	\includegraphics[width=.9\textwidth]{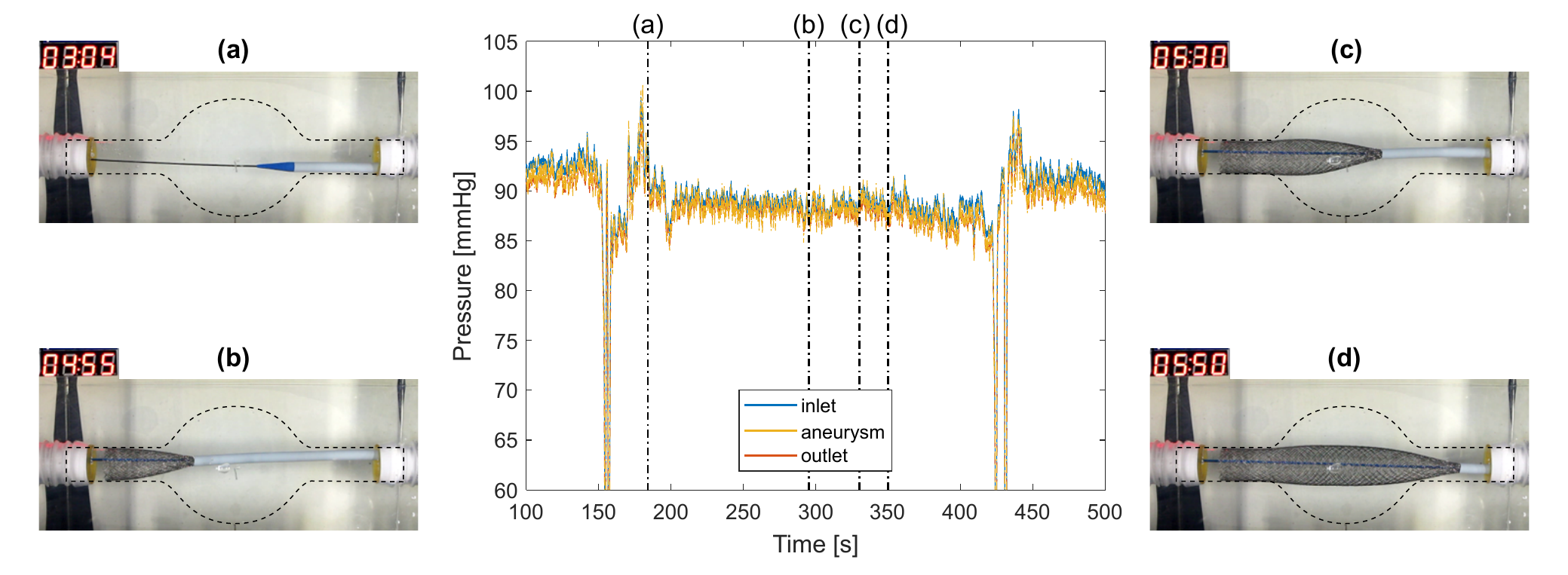}
	\caption{Evolution of aortic and aneurysmal pressure during MFM deployment in model 1 and frames of the recorded video at events of interest: (a) insertion of delivery system, (b) deployment reaches the aneurysm root, (c) deployment reaches the middle of the aneurysm and (d) deployment reaches the aneurysm end.}
	\label{fig:fig3}
\end{figure*}

\begin{figure*}[p]
	\centering
	\includegraphics[width=.9\textwidth]{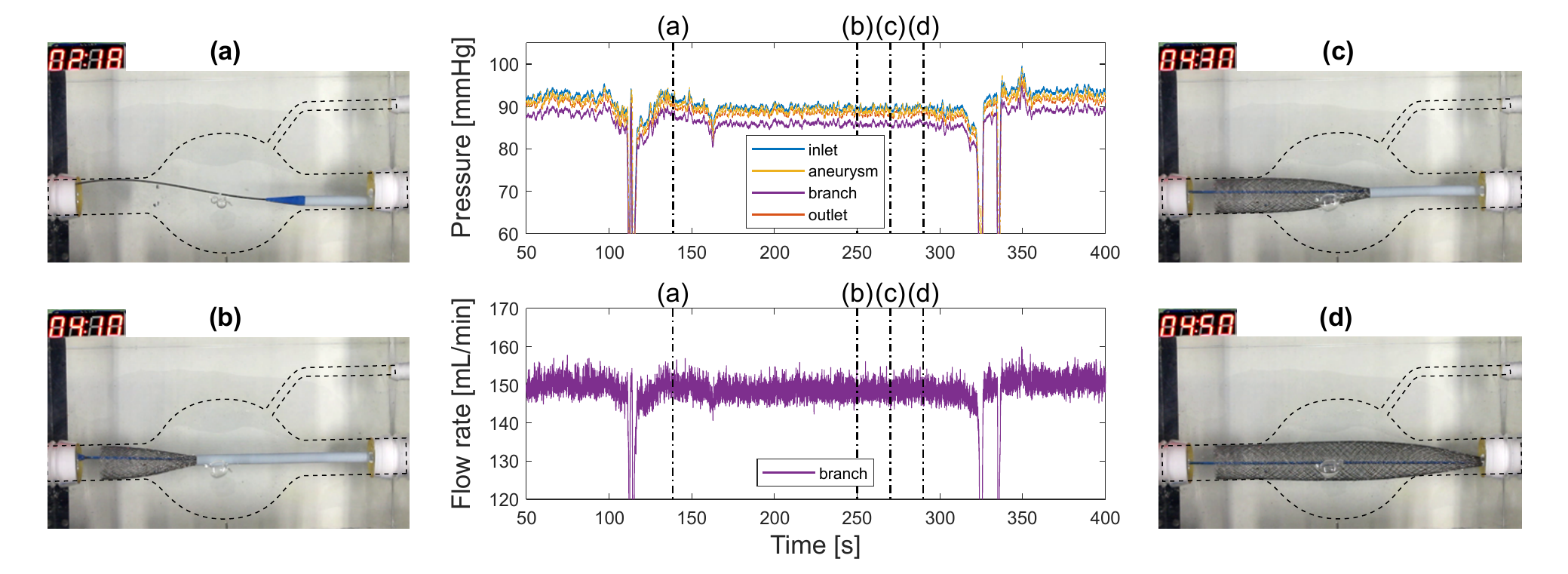}
	\caption{Evolution of pressure and flow rate during MFM deployment in model 2 and frames of the recorded video (events of interest as for figure 3).}
	\label{fig:fig4}
\end{figure*}

During this experiment, a qualified clinician (K.T.) deployed one MFM device inside each silicone model. The procedure lasted less than 5 minutes in both cases. Evolution of pressure and flow rate at each inlet/outlet and inside the aneurysm were monitored and correlated with videos of the deployment. They are presented in Figs.\ref{fig:fig3}-\ref{fig:fig4} along with snapshots of those videos. 

Opening of the introducer sheath for the insertion and removal of the delivery system led to a pressure drop in the system (visible at \num{150} and \SI{420}{\second} in Fig.\ref{fig:fig3}, \num{110} and \SI{330}{\second} in Fig.\ref{fig:fig4}). Hemodynamics perturbations were recorded at the beginning of the procedure due to the introduction of the guide-wire and delivery system (\SI{170}{\second} in Fig.\ref{fig:fig3}a). Pressure and flow rate evolutions then stabilized until the end of the deployment. The standard deviation of those evolutions was lower than \SI{10}{\percent} of the initial values.

When comparing the values before and after the procedure, results revealed no significant change of intra-aneurysmal pressure (Figs.\ref{fig:fig3}-\ref{fig:fig4}). In the model 2, the flow rate of the incorporated branch was also found unchanged (Fig.\ref{fig:fig4}). This result directly validates the preservation of incorporated branch perfusion after MFM device deployment.

\begin{table*}[p]
	\caption{Flow rate and pressure values measured during the PIV experiments. Values are reported as mean $\pm$ standard deviation. Pressure drop compared to inlet pressure is reported under bracket.}
	\vspace{1em}
	\centering
	\begin{tabular}{cccccccc}
		\toprule
		&	&	 \multicolumn{2}{c}{Flow rate [L/min]} & \multicolumn{4}{c}{Pressure [mmHg]} \\
		\cmidrule(r){3-4}\cmidrule(r){5-8}
		\multicolumn{2}{c}{Model}   & Inlet & Branch	& Inlet & Aneurysm & Branch & Outlet\\
		\midrule
		\multirow{2}{*}{AAA} & control 	& 3.0$\pm$.1	& ---			& 103$\pm$2 (0) & 103$\pm$2 (0)	& ---	& 102$\pm$2 (1)	\\
		& MFM 			& 3.0$\pm$.1	& ---			& 107$\pm$3 (0) & 106$\pm$3 (1)	& ---	& 106$\pm$3 (1)	\\
		\midrule
		\multirow{2}{*}{AAA with branch} & control & 3.1$\pm$.1	& .150$\pm$.003	& \enspace 98$\pm$2 (0) & \enspace 97$\pm$2 (1)	& \enspace 94$\pm$2 (4)	& \enspace 97$\pm$2 (1)	\\
		& MFM 			& 3.1$\pm$.1	& .149$\pm$.004	& 100$\pm$4 (0) & 100$\pm$4 (0) & \enspace 97$\pm$4 (3)	& \enspace 99$\pm$4 (1)	\\
		\bottomrule
	\end{tabular}
	\label{tab:table1}
\end{table*}

\subsection{Flow rate and pressure evaluation}

Flow rate and pressure values measured during the PIV experiments of each model are reported in Table~\ref{tab:table1}.

Experiments revealed no significant change of pressure or flow rate at all inlet/outlet positions or in the aneurysm with and without MFM device (Table~\ref{tab:table1}, \SI{<=5}{\percent}), in accordance to deployment experiments observations (Figs.\ref{fig:fig3}-\ref{fig:fig4}).

In all models, pressure difference between the aorta and the aneurysm was found low (\SI{<=1}{\milli\meter\mercury}) while the one between the aorta and branch, higher (\SI{>=3}{\milli\meter\mercury}). Those pressure differences were maintained after deployment of the MFM device. Moreover, branch flow rate was preserved after deployment of the MFM device.

\subsection{PIV experiments on AAA model}

\begin{figure*}[p]
	\centering
	\includegraphics[width=0.8\textwidth]{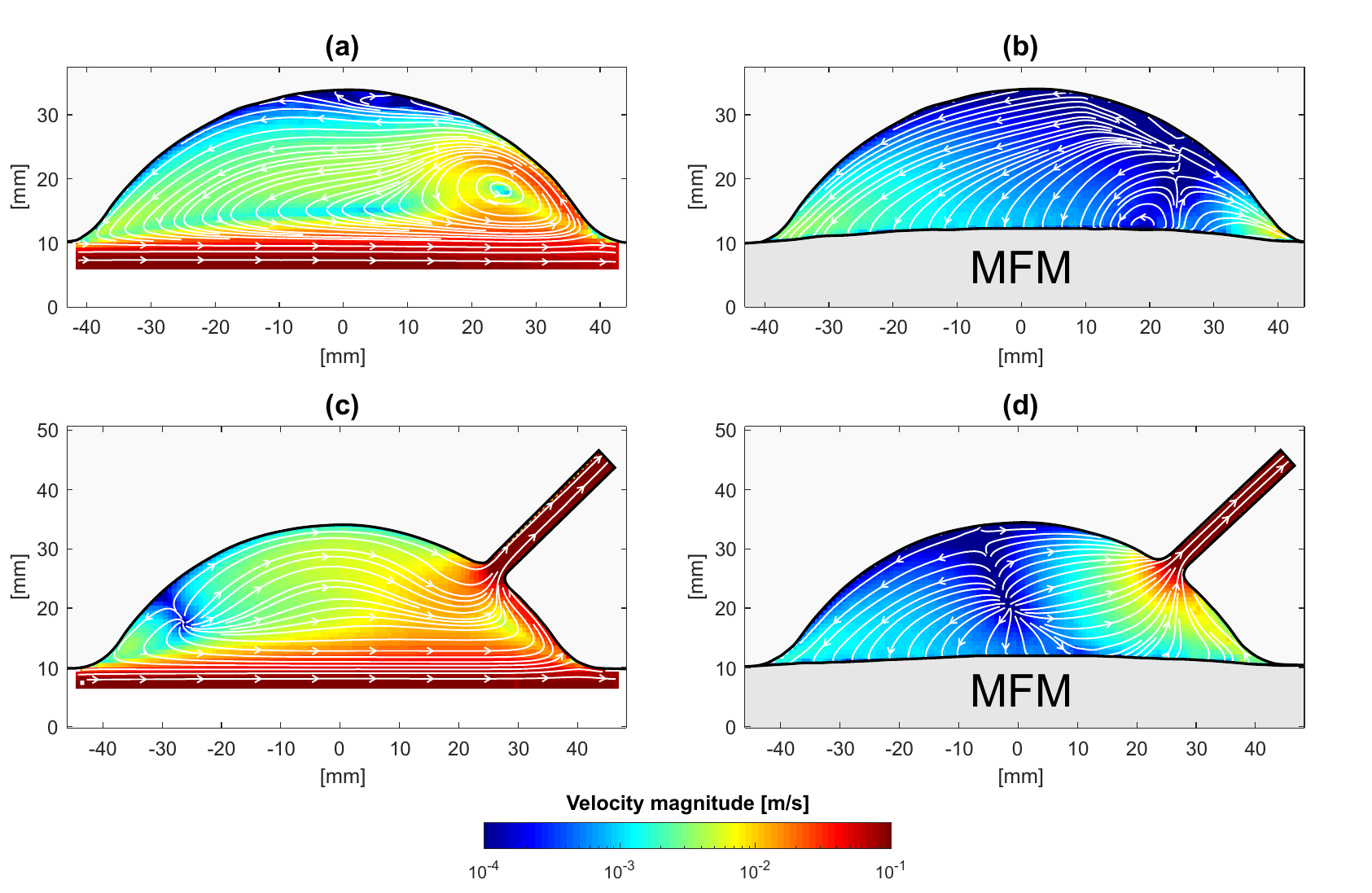}
	\caption{Velocity maps and streamlines evaluated by PIV in AAA model 1, (a) control and (b) MFM, and AAA model 2, (c) control and (d) MFM. Flow from left to right.}
	\label{fig:fig5}
\end{figure*}

\begin{figure*}[p]
	\centering
	\includegraphics[width=0.8\textwidth]{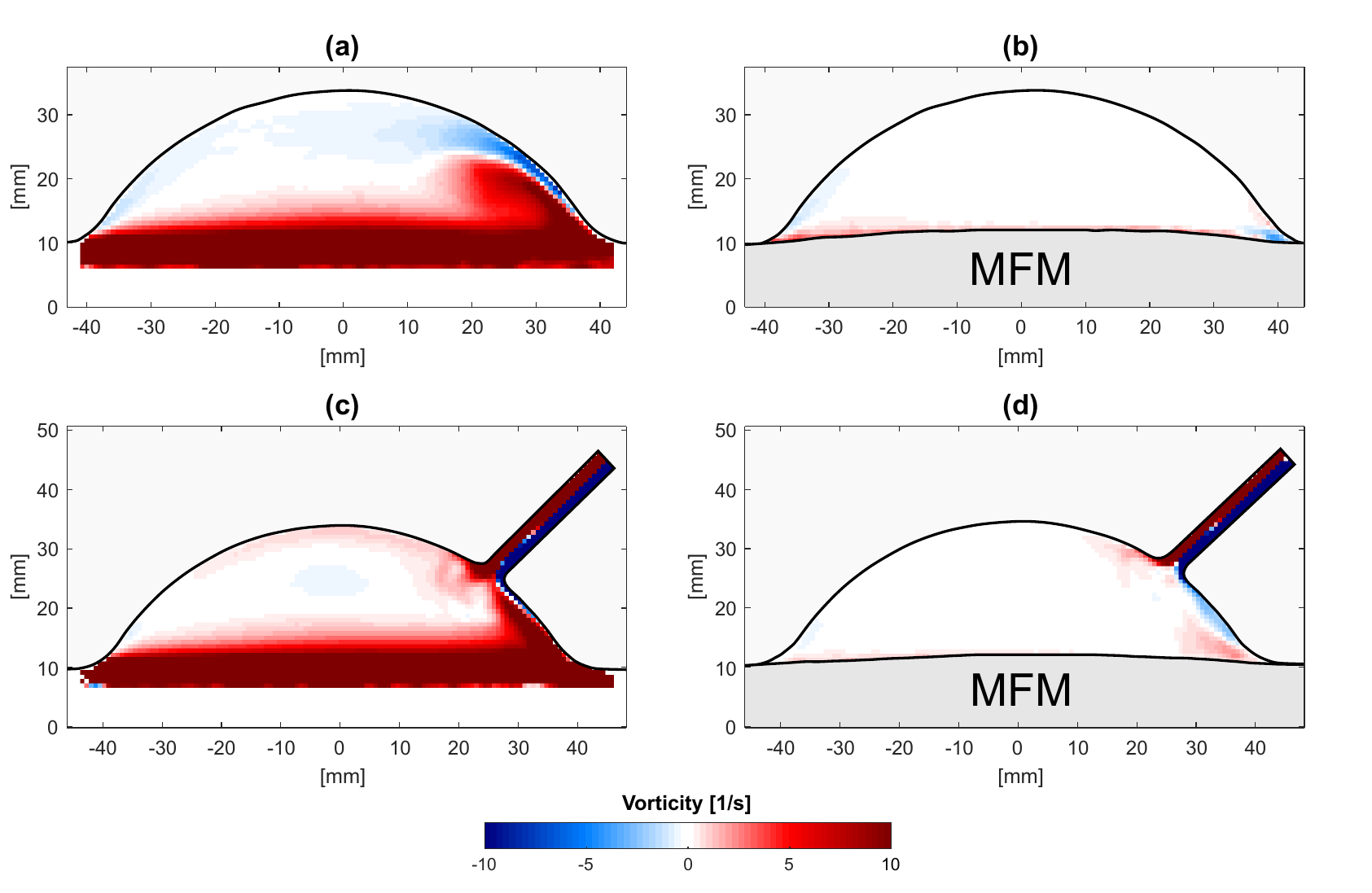}
	\caption{Vorticity maps evaluated by PIV in AAA model 1, (a) control and (b) MFM, and AAA model 2, (c) control and (d) MFM. Flow from left to right.}
	\label{fig:fig6}
\end{figure*}

Velocity and vorticity maps evaluated using PIV experiments are presented in Fig.\ref{fig:fig5} and Fig.\ref{fig:fig6}, respectively.

Results of PIV experiments conducted on the model 1 revealed significant changes in flow patterns and velocity in the aneurysm after deployment of the MFM device (Fig.\ref{fig:fig5}a-b). 

On the control model, a vortex is present close to the aneurysm end, inducing high velocity (Fig.\ref{fig:fig5}a) and vorticity (Fig.\ref{fig:fig6}a).

After deployment of the MFM, flow through the device is maintained. The vortex observed in the control model disappeared, leading to an important decrease of flow velocity (Fig.\ref{fig:fig5}b) and vorticity (Fig.\ref{fig:fig6}b), especially near the wall of the aneurysm end.

Mean velocity and enstrophy in the aneurysm decreased significantly from \SIrange[range-units = single]{5.43e-3}{7.05e-4}{\meter\per\second} and from \SIrange[range-units = single]{1.70e-2}{4.33e-4}{\square\meter\per\square\second}, respectively, before and after deployment of the MFM device. It corresponds to \SI{87.0}{\percent} and \SI{97.5}{\percent} decrease of average velocity and enstrophy, respectively.

\subsection{PIV experiments on AAA model incorporating a branch}

Results of PIV experiments conducted on the model 2 revealed significant changes in flow patterns and velocity in the aneurysm after deployment of the MFM device. 

On the control model 2, flow patterns were found different than in the model 1. The main flow is directed from the root to the branch (Fig.\ref{fig:fig5}c) with a twist near the aneurysm end, where velocity was found high. 

MFM device deployed in the model modified the flow patterns (Fig.\ref{fig:fig5}d). Flow directs straightly to the branch. A stagnation zone is observed near the wall at the center of the aneurysm. Flow velocity (Fig.\ref{fig:fig5}d) and vorticity (Fig.\ref{fig:fig6}d) were found greatly reduced in the aneurysm, while the flow velocity was maintained in the branch (Fig.\ref{fig:fig5}d).

Mean velocity and enstrophy in the aneurysm decreased significantly from \SIrange[range-units = single]{8.84e-3}{2.22e-3}{\meter\per\second} and from \SIrange[range-units = single]{2.03e-2}{7.49e-4}{\square\meter\per\square\second}, respectively, after deployment of the MFM device. It corresponds to \SI{74.9}{\percent} and \SI{96.3}{\percent} decrease of average velocity and enstrophy, respectively.

\section{Discussion}

To the best of our knowledge, this is the first experimental hemodynamics study on the MFM device. This study was made possible by a new resistance-based system capable of visually recording the deployment of the endovascular device while monitoring changes in pressure and flow rate. It also allowed further analysis of the intra-aneurysmal flow using PIV. This experimental system was found effective in simulating an endovascular procedure and studying the safety and effectiveness of endovascular devices. It can be easily adapted to analyze various endovascular procedures and combined with medical imaging to further reproduce clinical conditions.

\subsection{MFM device does not affect intra-aneurysmal pressure}

Using the developed experimental system, the deployment of a MFM by a trained radiologist was analyzed. Minimal hemodynamic changes were recorded during the deployment (Figs.\ref{fig:fig3}-\ref{fig:fig4}). Full patency of the branch and no critical increase or drop of pressure during the procedure validates the safety of the procedure.

After deployment, aneurysm and branch pressure were found similar to their initial values (Table~\ref{tab:table1}). Pressure drop between the aortic inlet and the aneurysm was also maintained.

Compared to usual EVAR covered device whose deployment induces intra-aneurysmal pressure decrease\parencite{19_Chong2003,24_Hinnen2007,25_Li2006}, uncovered braided stents does not affect this pressure. Several studies reported this behavior when using flow diverter, both \emph{in vivo}\parencite{26_Schneiders2011,27_Tateshima2014} and \emph{in silico}\parencite{28_Shobayashi2012,29_Wang2017}. 

\subsection{MFM device decreases intra-aneurysmal flow velocity and vorticity}

Intra-aneurysmal flow analysis performed by PIV revealed a strong decrease of flow velocity in the aneurysm both with and without incorporated branch. Moreover, vortex present before deployment disappeared thanks to the device, decreasing flow velocity and vorticity near the aneurysmal wall (Fig.\ref{fig:fig5}-\ref{fig:fig6}). Such conditions promote gradual thrombus formation from the aneurysmal wall, as seen \emph{in vivo}\parencite{30_Pinto2017,31_Finotello2019}.

Flow feeding the branch was modified by the MFM device in straight patterns (Fig.\ref{fig:fig5}d) with lower vorticity (Fig.\ref{fig:fig6}d). Intra-aneurysmal flow velocity exhibited a lower reduction (Fig.\ref{fig:fig5}b,d) while the branch perfusion is maintained (Table~\ref{tab:table1}). This effect might delay thrombus formation in favor of the perfusion preservation. The stagnation zone is found closer to the aneurysm root wall, where thrombus formation may start while branch perfusion persists.

\subsection{MFM device allows incorporated branch perfusion}

On the model with incorporated branch, the flow rate of the branch was maintained during the whole procedure (Fig.\ref{fig:fig4}) and confirmed equivalent after deployment (Fig.\ref{fig:fig5}c-d, Table~\ref{tab:table1}). This perfusion is made possible by the pressure difference between the aneurysm and the branch. As the MFM device does not modify intra-aneurysmal pressure, branch perfusion is maintained. This result is in agreement with animal\parencite{10_Sultan2015}, \emph{in vivo}\parencite{11_Benjelloun2016,32_Natrella_2012,33_Debing2014,34_Lowe2016,35_Kim2019,36_Euringer2011} and \emph{in silico}\parencite{37_Stefanov2016,38_Nezami2018} reports of branch flow perfusion preservation after MFM deployment. 

In the case of EVAR deployment, decrease of intra-aneurysmal pressure is observed\parencite{19_Chong2003,24_Hinnen2007,25_Li2006}. If this pressure becomes lower than the branch one, T2EL may occur\parencite{24_Hinnen2007,39_Sheehan2006}. T2EL is less likely to happen when using the MFM device as the branch pressure stays lower than the aneurysm one.

\subsection{Study limitations}
An idealized geometry and steady-state flow conditions were selected in this first experimental study to facilitate the analysis of flow patterns and to evaluate the main hemodynamic changes due to the MFM device. Patient-specific AAA geometry and pulsatile flow conditions will be applied in future experiments to further investigate the effect of the device.

\section{Conclusions}

Using a novel experimental system, the deployment of a MFM device in a AAA model was analyzed. The device did not change intra-aneurysmal pressure and preserved the incorporated branch perfusion during and after the procedure. PIV experiments revealed the ability of the device to decrease intra-aneurysmal velocity and vorticity. Preservation of the pressure drop between the aneurysm and the incorporated branch ensures branch patency and possible protection against endoleak type II. 

\section*{Funding}
This study was supported by JSPS KAKENHI (grant numbers JP18K18356 \& JP20H04557), ImPACT program of Council for Science, Technology and Innovation (Cabinet Office, Government of Japan) and Cardiatis (Isnes, Belgium).

\section*{Conflict of interest}
M.O. received a research grant from Cardiatis (Isnes, Belgium).

\normalsize
\printbibliography

\end{document}